\documentclass[useAMS,usenatbib]{mn2e}

\usepackage{graphicx}
\usepackage{natbib}
\usepackage[figuresleft]{rotating}
\newcommand{\HI}{H{\,\small I}}

\newcommand{\ltsima} {$\; \buildrel < \over \sim \;$}
\newcommand{\gtsima} {$\; \buildrel > \over \sim \;$}
\newcommand{\lta} {\lower.5ex\hbox{\ltsima}}
\newcommand{\gta} {\lower.5ex\hbox{\gtsima}}
\newcommand{\kmsMp}{km s$^{-1}$\,Mpc$^{-1}$}
\newcommand{\kms}{km\ s$^{-1}$}
\newcommand{\FRI}{FR{-\small I}}
\newcommand{\FRII}{FR{-\small II}}

\title[Radio source propagating into H\,I disc in NGC~3801]{Classical radio source propagating into outer H\,I disc in NGC~3801}
\author[B. H. C. Emonts et al.]{B. H. C. Emonts$^{1}$\thanks{E-mail:bjorn.emonts@csiro.au}, C. Burnett$^{2,1}$\thanks{ATNF Summer Student 2009/10}, R. Morganti$^{3,4}$, C. Struve$^{3,4}$\\
$^{1}$CSIRO Astronomy and Space Science, Australia Telescope National Facility, PO Box 76, Epping NSW, 1710, Australia\\
$^{2}$School of Physics, University of Melbourne, Parkville, Victoria 3010, Australia\\
$^{3}$Netherlands Institute for Radio Astronomy, Postbus 2, 7990 AA Dwingeloo, the Netherlands\\
$^{4}$Kapteyn Astronomical Institute, University of Groningen, P.O. Box 800, 9700 AV Groningen, the Netherlands
}
\begin{document}

\date{}

\pagerange{\pageref{firstpage}--\pageref{lastpage}} \pubyear{2010}

\maketitle

\label{firstpage}

\begin{abstract}
We present observations of a large-scale disc of neutral hydrogen (\HI) in the nearby Fanaroff $\&$ Riley type-I radio galaxy NGC~3801 with the Westerbork Synthesis Radio Telescope. The \HI\ disc (34 kpc in diameter and with $M_{\rm HI} = 1.3 \times 10^{9} M_{\odot}$) is aligned with the radio jet axis. This makes NGC~3801 an ideal system for investigating the evolution of a small radio source through its host galaxy's cold ISM. The large-scale \HI\ disc is perpendicular to a known inner CO disc and dust-lane. We argue that the formation history of the large-scale \HI\ disc is in agreement with earlier speculation that NGC~3801 was involved in a past gas-rich galaxy-galaxy merger (although other formation histories are discussed). The fact that NGC~3801 is located in an environment of several \HI-rich companions, and shows indications of ongoing interaction with the nearby companion NGC~3802, strengthens this possibility. The large amounts of ambient cold ISM, combined with X-ray results by \citet*{cro07} on the presence of over-pressured radio jets and evidence for an obscuring torus, are properties that are generally not, or no longer, associated with more evolved \FRI\ radio sources. We do show, however, that the \HI\ properties of NGC~3801 are comparable to those of a significant fraction of nearby low-power compact radio sources, suggesting that studies of NGC~3801 may reveal important insight into a more general phase in the evolution of at least a significant fraction of nearby radio galaxies.
\end{abstract}


\section{Introduction}
\label{sec:intro}
Over the past decade it has become clear that radio sources emanating from super-massive black-holes in the centres of galaxies are - rather than merely a by-product of galaxy evolution - key players in the formation and evolution of galaxies throughout the Universe. It is now fairly well established that the properties of Active Galactic Nuclei (AGN) are intimately linked with the properties of their host galaxies.
While empirical evidence has established the co-evolution of massive black holes and their host galaxies at various redshifts (e.g. \citealt*{fer00}, \citealt{ale05}, see also \citealt*{sil98}), numerical simulations demonstrate that AGN-induced outflows are important feedback processes for clearing the circum-nuclear regions and halting the growth of the super-massive black-holes, regulating the correlations found between the black-hole mass and the host galaxy's bulge properties and preventing the formation of too many massive galaxies in the early Universe (e.g. \citealt*{dim05}, \citealt*{spr05}, \citealt{hop05}, \citealt{wag11}).

An important phase of AGN feedback is when synchrotron jets emanating from the central black-hole region start propagating into the host galaxy's inter-stellar medium (ISM). At high redshifts, radio sources are observed to be aligned with optical continuum from warm gas and stars (\citealt*{cha87}, \citealt{mcc87}) and the gas in these regions is characterised by highly disturbed kinematics (\citealt*{mcc96}, \citealt*{vil99}, \citealt{hum06}). This indicates that radio sources vigorously interact with the host galaxy's ISM, entrain and expel gas and induce or quench star formation \citep[e.g.][]{bic00}. At intermediate and low redshifts, powerful radio jets are know to induce fast ($>1000$\,\kms) outflows of ionised and neutral gas \citep[e.g.][]{tad91,cla98,oos00,mor03,mor05a,mor05,mor07,emo05,hol06,hol08,hol11} and can shock-heat the molecular gas \citep{gui11}, exerting significant feedback on the host galaxy's ISM.

While propagating radio jets thus have a profound impact on their surrounding ISM, the reverse is also true \citep[see simulations by e.g.][]{sax02b,sax02a,jey05,bic06,sut07}. It is well established that external factors, in particular the inter-stellar and inter-galactic medium (ISM and IGM), shape the morphological and physical properties of radio sources. For example, there is strong evidence that radio sources in general show a degree of intrinsic asymmetry (apart from orientation effects), which is most prominent in compact sources \citep[][]{sai95,sai01,sai03a}. Compact radio sources also show a larger degree of polarisation asymmetry than extended sources, suggesting a link between the propagating radio jets and surrounding ISM \citep*{sai03b}. Also, the impact of supersonic jets onto the ISM/IGM can create cocoons of (shocked) gas and synchrotron emission with `hot-spot' morphologies, while entrainment of gas by slower jets can lead to bright radio sources with edge-darkened lobes \citep[see e.g. reviews by][]{bri84,fer98,bic06}. These external factors may thus contribute to the classical dichotomy in radio sources between Fanaroff $\&$ Riley type-I  \citep[\FRI;][]{fan74} objects, with low-power, sub-relativistic jets that have an edge-darkened morphology, and type-II (\FRII) sources, with powerful, relativistic jets ending in a bright hot-spot \citep[see e.g.][]{gop00}. 


A particularly interesting object in this respect is the nearby ($z = 0.0115$) radio galaxy NGC~3801. NGC~3801 is classified as an \FRI\ radio galaxy, based on its total radio power \citep[$P_{\rm 1.4\,GHz} = 2.9 \times 10^{23}$\,W\,Hz$^{-1}$;][]{con88}, radio source morphology \citep{jen82,cro07} and weak emission-line spectrum \citep{hec86}. The radio source of NGC~3801 is small \citep[$\sim 11$ kpc;][]{jen82} and contained well within the optical boundaries of a peculiar early-type host galaxy, with a patchy dust feature along the major stellar axis and a perpendicular prominent inner dust-lane \citep{hec86,ver99}, as well as a distorted and box-shaped faint outer envelope \citep{hec86}. The inner dust-lane coincides with a kpc-scale CO disc, oriented perpendicular to the radio axis (\citealt{das05}, see also \citealt{oca10}). 

X-ray observations by \citet{cro07} showed evidence for shocked X-ray gas in the central region of NGC~3801, created by supersonic, over-pressured jets that are not in equilibrium with the hot ISM. These X-ray studies also revealed a high absorbing column towards the nucleus of NGC~3801, likely due to an obscuring torus \citep{cro07}. In addition, single-dish observations showed that NGC~3801 is rich in neutral hydrogen (\HI) gas \citep{hec83,dup96}. Based on its cold gas content and peculiar optical host galaxy morphology, it has been suggested that NGC~3801 could have been involved in a gas-rich galaxy-galaxy merger \citep{das05,cro07}. As speculated by \citet{cro07}, these are all properties that are generally not associated with \FRI\ sources, but are believed to be common among more powerful, high-excitation radio-loud AGN (see also \citealt{hec86}, \citealt{bau92}, \citealt*{har06}, \citealt{cro08}, as well as Sect. \ref{sec:classical}).

In this paper we present radio synthesis observations of the \HI\ gas in NGC~3801. Our observations reveal that the \HI\ gas is distributed in a large-scale disc of cold gas. The large-scale \HI\ disc is aligned with the radio axis, indicating that the radio jets propagate directly into the cold gas disc. Based on our \HI\ results, we discuss the formation history of NGC~3801 and investigate further how unique NGC~3801 is compared to other nearby \FRI\ radio galaxies.

Throughout this paper we assume H$_{0} = 71$\,\kmsMp, which puts NGC~3801 at a distance of 48.6 Mpc and 1 arcsec = 0.24 kpc.

\section{Observations}
\label{sec:observations}

\HI\ observations of NGC~3801 were performed with the Westerbork Synthesis Radio Telescope (WSRT) on 13 July 2009 during a single 12h track. A standard calibration was applied, using 3C\,147 as primary calibrator. Data were reduced and analysed using the {\sc MIRIAD} and {\sc KARMA} software. After calibration, the continuum (`channel-0') data were separated from the line data by fitting a straight line to the line-free channels. A radio continuum map with uniform weighting and beam-size $10.57'' \times 37.55''$ (PA -0.4$^{\circ}$) was produced by Fourier transforming the channel-0 data and subsequently cleaning the signal. In the same way, an \HI\ cube with robust weighting +1 \citep{bri95}, beam-size of $22.47'' \times 87.52''$ (PA 0.2$^{\circ}$) and channel separation of 4.1 \kms\ was created from the line-data.\footnote{The large difference in beam-size between north-south and east-west direction is a result of the relatively low elevation of the source throughout the observations with the east-west array-configuration of the WSRT.} We subsequently Hanning smoothed the line-data to an effective velocity resolution of $\sim 8$ \kms, resulting in a noise level of 0.7 mJy\,bm$^{-1}$\,chan$^{-1}$. Total intensity images of the \HI\ emission were made by using a mask (derived by smoothing the positive signal both spatially and kinematically) to extract the \HI\ line emission from the original line data-set. The data in this paper are corrected for primary beam attenuation and presented in optical barycentric velocity definition.

\section{Results}
\label{sec:results}






\begin{figure*}
\centering
\includegraphics[width=\textwidth]{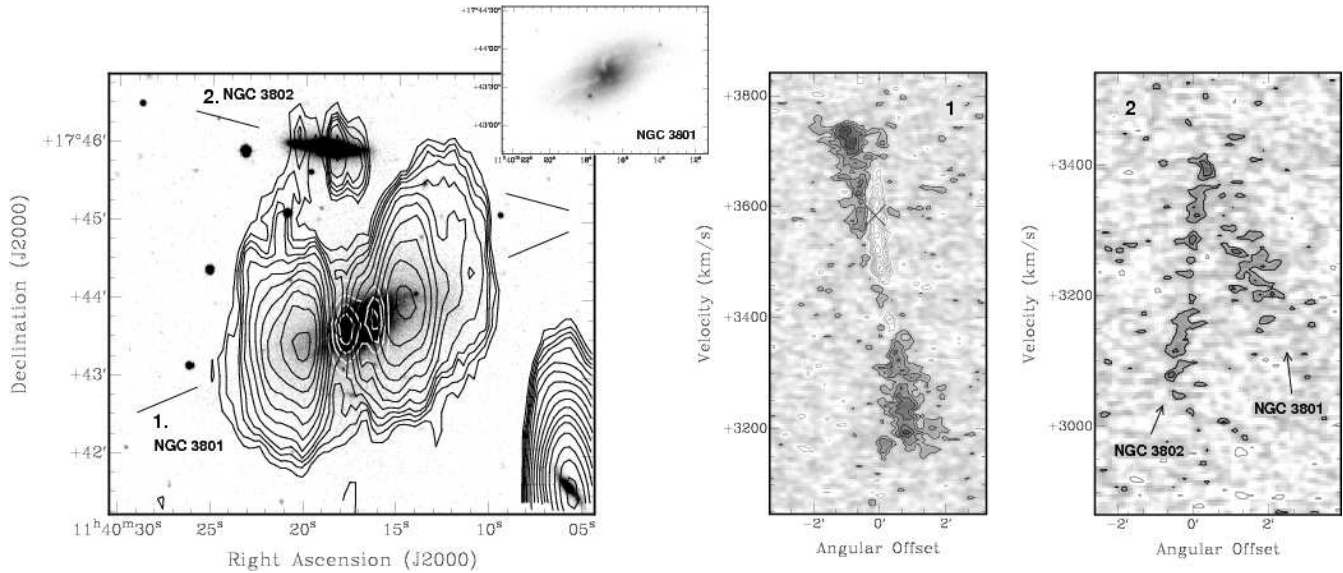}
\caption{NGC~3801 and NGC~3802. {\sl Left:} Total intensity of \HI\ emission (black contours) and 1.4\,GHz radio continuum (white contours) overlaid into an optical SDSS image of NGC~3801 (see also top-right inset) and NGC~3802. Note that the spatial resolution of the radio data in north-south direction is a factor of 4 worse than in east-west direction due to the highly elongated beam (Sect. \ref{sec:observations}). For clarity, \HI\ absorption has been omitted from this plot. Contour levels \HI: 0.15, 0.21, 0.28, 0.34, 0.45, 0.62, 0.84, 1.1, 1.5, 2.0, 2.5, 3.1, 3.8, 4.5, 5.2, 6.0, 7.0, 8.0 $\times 10^{20}$\,cm$^{-2}$. Contours levels radio continuum: 0.04, 0.11, 0.18, 0.25 Jy\,bm$^{-1}$. {\sl Right:} Position velocity maps of the \HI\ emission (black contours) and \HI\ absorption (grey contours) along the two lines presented in the left image. Contour levels: -9.0, -7.2, -5.6, -4.2, -3.0, -2.0, -1.2 (grey), 1.2, 2.0, 3.0, 4.2 (black) mJy\,bm$^{-1}$. The cross represents the peak of CO emission from `clump C' found by \citet{das05} (see Sect. \ref{sec:discussiondisc}).}
\label{fig:HIdisc}
\end{figure*}

\begin{figure*}
\centering
\includegraphics[width=\textwidth]{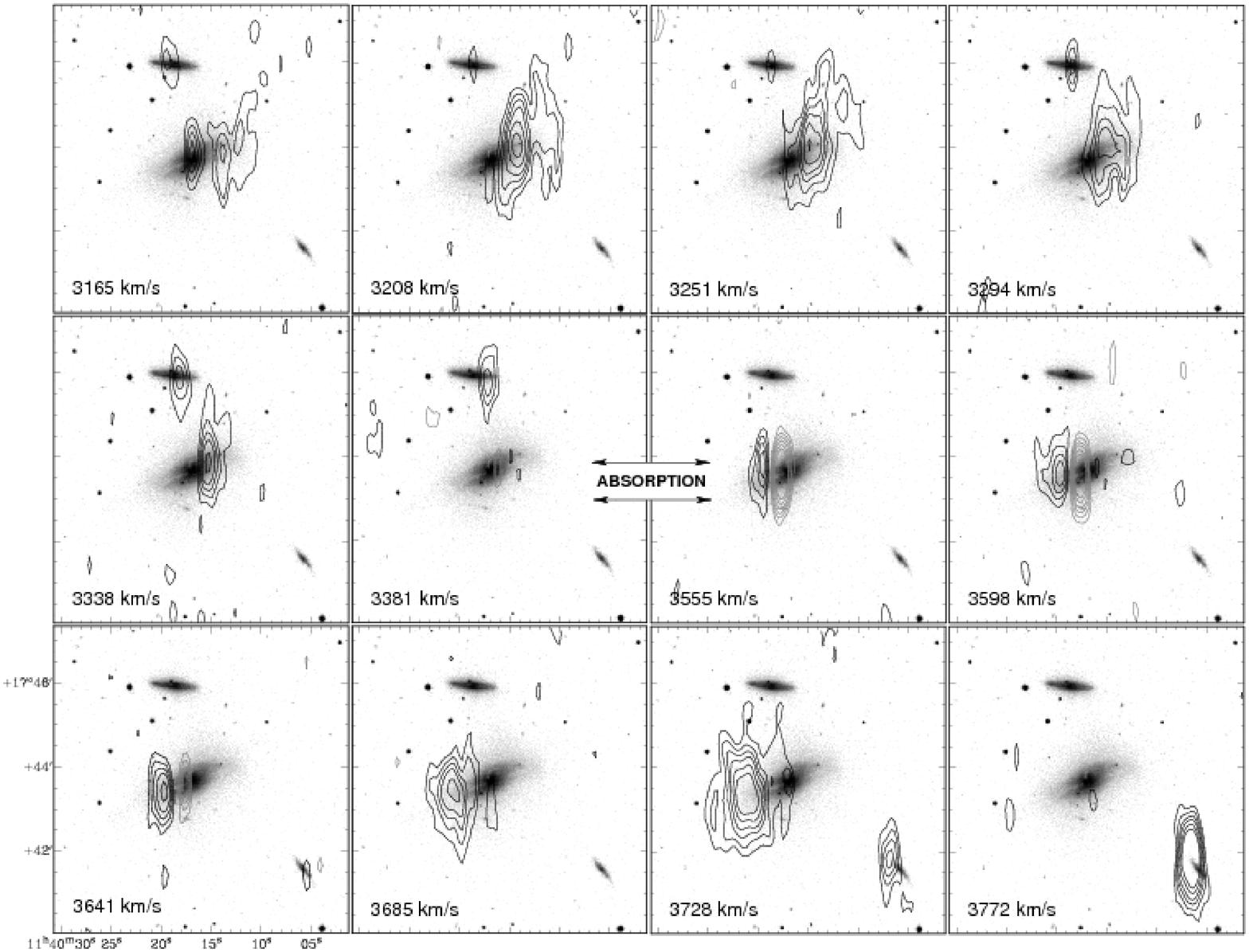}
\caption{Channel maps of the \HI\ emission in NGC~3801, binned to a velocity resolution of 43 \kms. The velocity range of the \HI\ absorption is largely omitted from this plot. Contour levels: -3.2, -2.6, -2.0, -1.6, -1.2, -0.8 (grey), 0.8, 1.2, 1.6, 2.0, 2.6, 3.2 (black) mJy\,bm$^{-1}$.}
\label{fig:HIchanmaps}
\end{figure*}

\subsection{\HI\ emission in NGC~3801/3802}
\label{sec:disc}

Figure \ref{fig:HIdisc} show the large-scale \HI\ disc of NGC~3801, which has a diameter of 34~kpc. A small fraction of the \HI\ emission-line gas in the disc was detected with VLA observations by \citet{hot09}. We derive a total mass of the emission-line gas of $M_{\rm HI} = 1.3 \times 10^{9} M_{\odot}$, though we note that part of the \HI\ disc in our WSRT observations is detected in absorption against the radio continuum source (see Fig. \ref{fig:absorption}) and not taken into account in this \HI\ mass estimate. The \HI\ emission-line gas is aligned with the host galaxy's main stellar axis and the east-west dust-lane. The \HI\ disc shows regular rotation with a large velocity spread of $\Delta v \approx 600$ \kms, corresponding to a maximum rotational velocity of $\sim 300$ \kms\ \citep[which is larger than the rotational velocity of the optical emission-line gas;][]{hec83,hec85spec}. The \HI\ disc is centred on v$_{\rm sys} = 3452 \pm 20$ \kms\ ($z = 0.01151 \pm 0.00007$), which we argue reflects the systemic velocity of NGC~3801.
Despite the regular rotation, Fig. \ref{fig:HIchanmaps} shows that the \HI\ disc is not (yet) fully settled. The most prominent asymmetry in the \HI\ distribution occurs at the north-western edge of the disc, with gas apparently stretching in the direction of the close gas-rich companion NGC~3802. 

The close companion NGC~3802 is a disc-galaxy (S0 or spiral) approximately 33 kpc north of NGC~3801. NGC~3802 contains $M_{\rm HI} \approx 3.9 \times 10^{7} M_{\odot}$ of \HI\ gas. The \HI\ gas has an asymmetric distribution across NGC~3802, peaking in the western part of the disc and continuing on the eastern side, covering a total velocity range of $\sim$\,400 \kms\ (see Fig. \ref{fig:HIdisc} {\sl - right} for more details). While the total intensity image of Fig. \ref{fig:HIdisc} {\sl (left)} shows a spurious \HI\ bridge in between NGC~3801 and the {\sl eastern} part of the disc of NGC~3802, we note that this is merely a projection effect. In fact, Figs. \ref{fig:HIdisc} {\sl (right)} and \ref{fig:HIchanmaps} show that there is an apparent bridge or warp of \HI\ gas stretching from the north-western part of the \HI\ disc of NGC~3801 towards the {\sl western} part of the disc of NGC~3802. This \HI\ feature is too faint to be identified in the total intensity image of Fig. \ref{fig:HIdisc} {\sl (left)}. The asymmetric \HI\ distribution in companion galaxy NGC~3802 and the apparent \HI\ bridge in between NGC~3801 and NGC~3802 suggests that the NGC~3801/3802 system is in ongoing interaction.


\subsection{\HI\ absorption in NGC~3801}
\label{sec:absorption}

\begin{figure*}
\centering
\includegraphics[width=\textwidth]{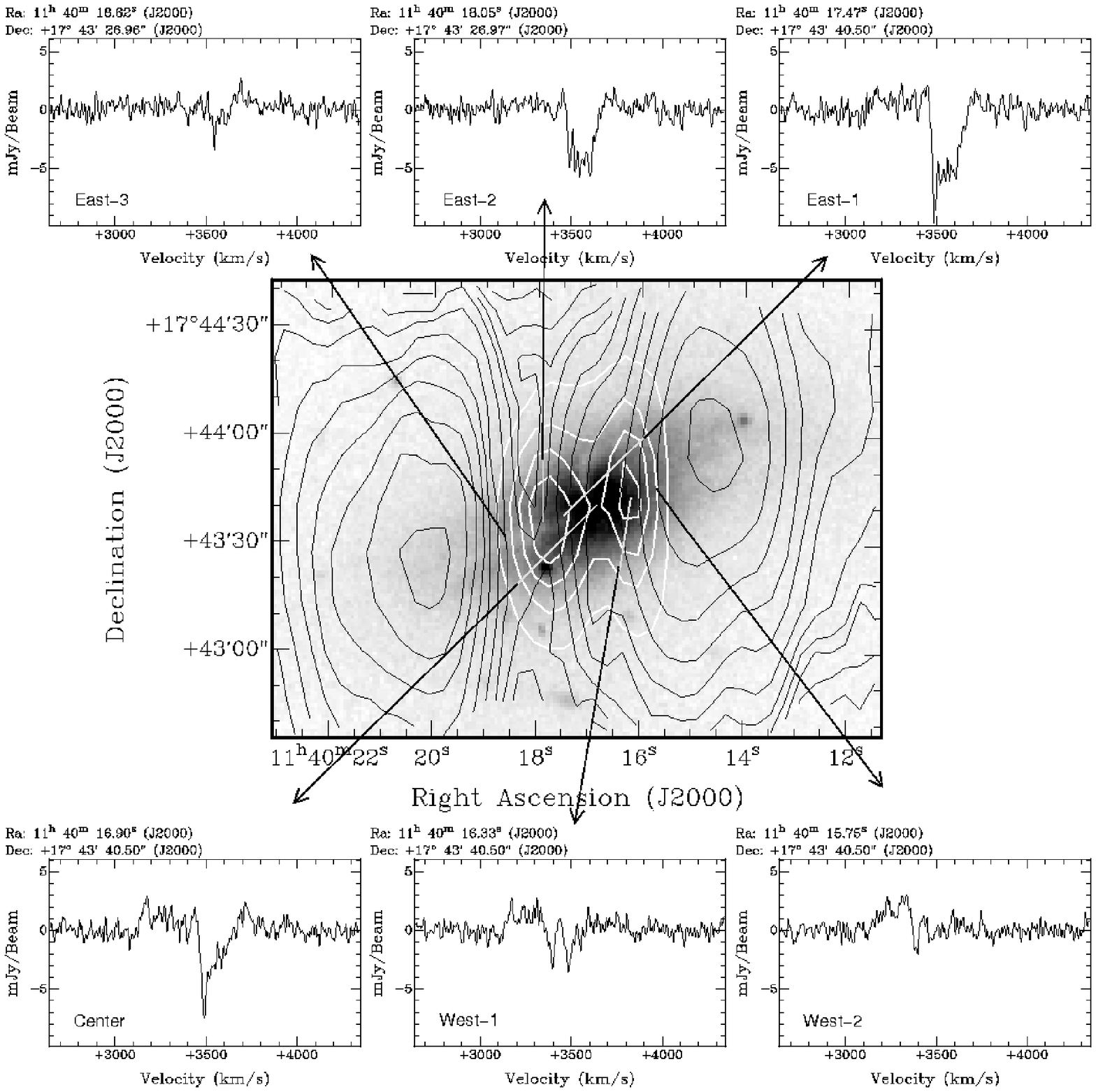}
\caption{\HI\ absorption properties of NGC~3801. The middle plot shows a zoom-in of the optical host galaxy, \HI\ emission and radio continuum from Fig. \ref{fig:HIdisc}. The zoom-windows show the \HI\ profile at various locations along the \HI\ disc.}
\label{fig:absorption}
\end{figure*}

\begin{figure*}
\centering
\includegraphics[width=\textwidth]{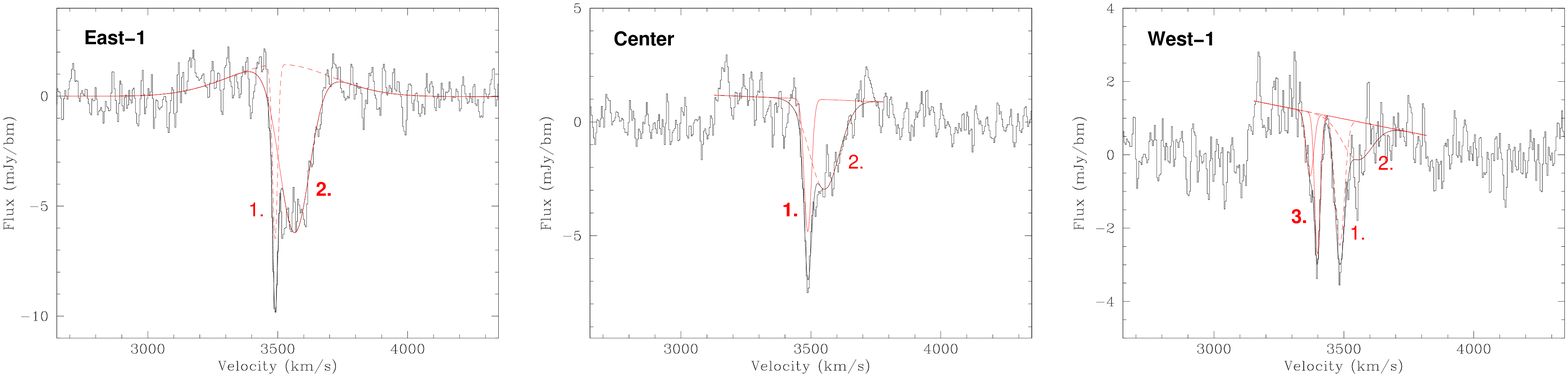}
\caption{Gaussian fits to the \HI\ absorption profiles of Fig. \ref{fig:absorption}. The black solid line shows the total fit, while the red lines show the individual Gaussian components. The numbers correspond to the absorption features as listed in Table \ref{tab:absorption}. The actual values in Table \ref{tab:absorption} are derived from the Gaussian profiles with solid red lines and thick numbers in this plot.}
\label{fig:fit}
\end{figure*}

Figure \ref{fig:absorption} shows the \HI\ absorption detected against the radio continuum source in the central part of NGC~3801. For the radio continuum we derive a total integrated flux of $S_{\rm 1.4\,GHz} = 1.03$\,Jy, corresponding to $P_{\rm 1.4\,GHz} = 2.9 \times 10^{23}$\,W\,Hz$^{-1}$. This is in agreement with earlier results by \citet{jen82} and \citet{con88}.

Taking into account that the \HI\ absorption spectra shown in Fig. \ref{fig:absorption} are not entirely mutually independent, there are three distinctive features visible. These features are visualised in Fig. \ref{fig:fit} and summarised in Table \ref{tab:absorption}. 

The first is a deep and narrow \HI\ absorption at $v = 3488 \pm 20$ \kms\ ($z = 0.01163 \pm 0.00007$), which peaks at the location of the nucleus. The central velocity of this feature is in good agreement with the central velocity of $v = 3494 \pm 70$ \kms\ that \citet{weg03} derive from optical spectroscopy.

The second is a broad component, covering the central region and eastern radio lobe. Figure \ref{fig:HIdisc} {\sl (middle)} suggests that this broad absorption component traces \HI\ gas as part of the large-scale \HI\ disc, at locations where the gas is in front of the radio source.\footnote{We note that the velocity gradient of the \HI\ absorption as seen in Fig. \ref{fig:HIdisc} could be slightly steeper than that of the \HI\ emission. This may occur because the spatial resolution of our observations is coarser than the underlying radio continuum structure, hence the \HI\ absorption is less affected by beam-dilution than \HI\ emission and could trace the more inner regions of the \HI\ disc.} This also means that the eastern radio jet is most likely the `far' side of the radio source (where the radio continuum is behind the absorbing gas in the disc), while the western jet is the `near' side.  

The third feature is a narrow absorption against the western jet, which is blueshifted by $\sim$\,50 \kms\ with respect to v$_{\rm sys}$. Fig. \ref{fig:HIdisc} {\sl (middle)} shows that also this \HI\ component is most likely part of the large-scale, edge-on \HI\ disc that embeds the small radio source, although it cannot be ruled out that this blueshifted \HI\ absorption represents an outflow of neutral gas driven by the propagating radio jets. Figure \ref{fig:absorption} also shows indications of a narrow and much weaker \HI\ absorption component against the eastern radio jet in NGC~3801, around v\,$\sim$\,3550 \kms.

Higher resolution \HI\ absorption observations are essential for studying the nature of the various absorption features in more detail. For this, we refer to ongoing work by \citet{hot09}.

\subsection{\HI\ environment}
\label{sec:companions}

\begin{figure*}
\centering
\includegraphics[width=\textwidth]{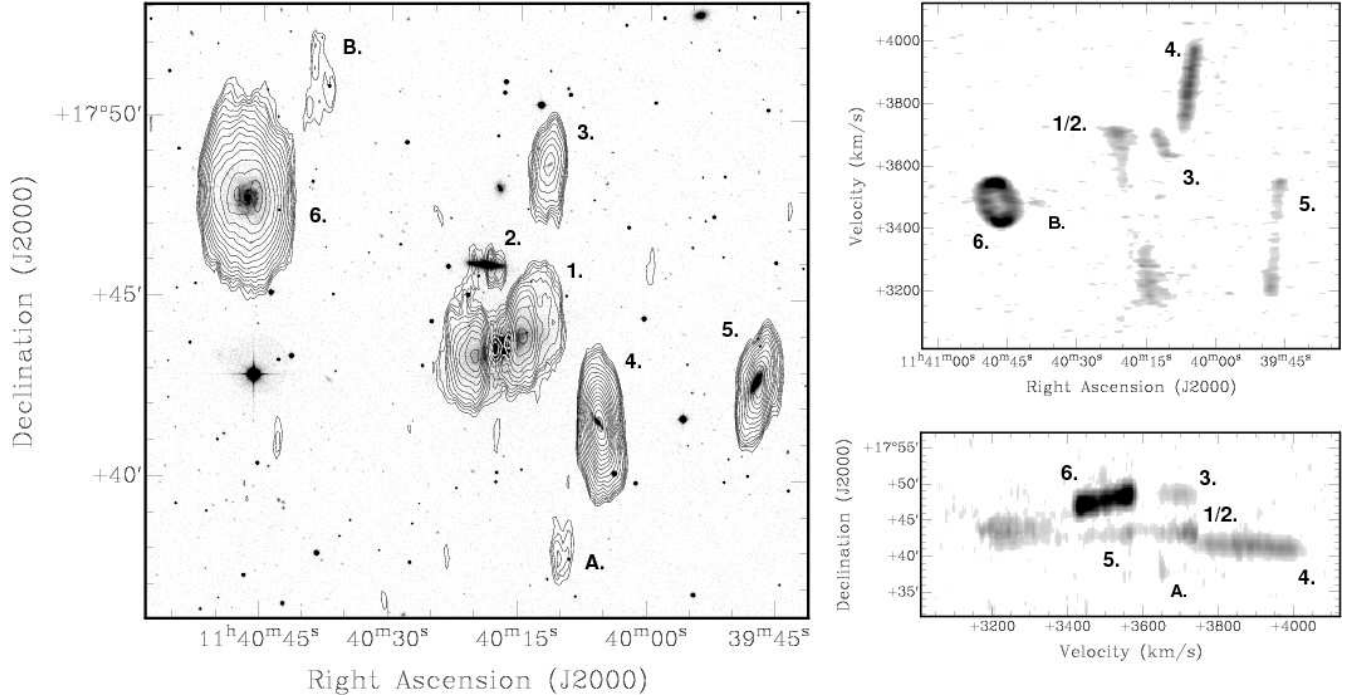}
\caption{\HI\ environment of NGC~3801. {\sl Left:} Contours of the total intensity \HI\ emission (black) overlaid onto an optical SDSS image. Contour levels \HI: 0.15, 0.21, 0.28, 0.34, 0.45, 0.62, 0.84, 1.1, 1.5, 2.0, 2.5, 3.1, 3.8, 4.5, 5.2, 6.0, 7.0, 8.0 $\times 10^{20}$\,cm$^{-2}$. The 1.4\,GHz radio continuum source is represented by white contours in the very central region of NGC~3801 (levels: 0.04, 0.11, 0.18, 0.25 Jy\,bm$^{-1}$). For clarity, \HI\ absorption has been omitted from this plot. {\sl Right:} Position-velocity plots of the same \HI\ data cube, represented as total intensity of the \HI\ emission integrated along the RA and dec axes.}
\label{fig:HIenvironment}
\end{figure*}

Figure \ref{fig:HIenvironment} shows the \HI\ environment of NGC~3801, with five other galaxies detected in \HI\ emission. Table \ref{tab:companions} summarises the \HI\ properties of these systems. The bulk of the \HI\ in these companion galaxies is distributed in disc-like structures. 

We also find two `clouds' of \HI\ gas that are not associated with an optical counterpart in the SDSS image of Fig. \ref{fig:HIenvironment}. One of them, cloud A, has a velocity that coincides with that of the eastern part of the \HI\ disc of NGC~3801, as well as companion 3, and is located 83 kpc south of NGC~3801 (note that the velocity of cloud A is significantly offset from that of the much nearer companion 4; see Fig. \ref{fig:HIenvironment} -- {\sl left}). Cloud B coincides in both velocity and location with the \HI-rich companion 6 (NGC~3806) and is most likely gas that is either stripped off, or being accreted onto, the \HI\ disc of NGC~3806. If these clouds of \HI\ gas are tidal in origin, and given that the total \HI\ mass of both clouds ($4-5 \times 10^{7}$ M$_{\odot}$) is similar to that of NGC~3801's close companion NGC~3802, this indicates that past as well as ongoing interactions could have a profound influence on the evolution of this group of galaxies.

\begin{table}
\caption{\HI\ absorption properties}
\vspace{0.4cm}
\label{tab:absorption}
\begin{tabular}{cccccc}
Feature & v$_{\rm centre}$ & FWHM & $\tau$ & N$_{\rm HI}$ \\
$\#$ & (\kms) & (\kms) & & ($\times 10^{20}$\,cm$^{-2}$) \\
\hline
1 & 3488 & 33  & 0.035 & 2.1 \\
2 & 3564 & 136 & 0.028 & 7.0 \\
3 & 3399 & 42$^{*}$  & 0.014 & 1.1$^{*}$ \\
\end{tabular} 
\flushleft 
{Notes -- Values have been derived from the Gaussian fits to the profiles, as shown in Fig. \ref{fig:fit}. v$_{\rm centre}$ is the central velocity of the absorption feature (error $\pm 20$\,\kms); FWHM is the full width at half the maximum intensity; $\tau$ is the maximum optical depth that the feature displays across the radio continuum and is calculated through \(e^{-\tau}=1-{{S_{\rm abs}}\over{S_{\rm cont}}}\) (with $S_{\rm cont}$ the underlying 21cm radio continuum flux at the limit spatial resolution of our observations); $N_{\rm HI}$ is the derived \HI\ column density (assuming $T_{\rm spin}$ = 100\,K).\\
$^{*}$derived from both Gaussian components that comprise feature 3 in Fig. \ref{fig:fit}.}
\end{table} 

\section{Discussion}
\label{sec:discussion}

\subsection{Large-scale \HI\ disc}
\label{sec:discussiondisc}

Our \HI\ results show that the bulk of the \HI\ gas detected in earlier single-dish observations \citep{hec83,dup96} is arranged in an \HI\ disc with a diameter of 34 kpc, aligned along the host galaxy's major stellar axis and prominent east-west dust-lane. We therefore argue that this large-scale \HI\ disc traces the major axis of the host galaxy. The peak \HI\ surface density in the disc is $\Sigma_{\rm HI} = 2.1$ M$_{\odot}$\,pc$^{-2}$. This is below (or at best close to) the critical \HI\ surface density for star formation typically found in nearby galaxies \citep{ken89,hul93,mar01,big08,ler08}, as also predicted by models \citep[e.g.][]{sch04}. It is therefore unlikely that wide-spread star formation is happening in the disc (although star formation may occur in regions of higher local density).

The rotation of the large-scale \HI\ disc allows us to estimate the dynamical mass of NGC~3801 to be M$_{\rm dyn} = ({\rm Rv}^{2}/{\rm G}) \cdot {\rm sin^{-2}}i = 3.6 \times 10^{11}$\,M$_{\odot}$ (with R\,=\,17\,kpc the radius of the \HI\ disc, v\,=\,300\,\kms\ its rotation velocity and assuming the \HI\ disc is viewed edge-on, i.e. $i = 90^{\circ}$). Given that $L_{\rm B} = 2.3 \times 10^{10}L_{\odot}$ \citep*[][corrected for H$_{0} = 71$\,\kmsMp]{hec85b}, NGC~3801 has M$_{\rm dyn}$/$L_{\rm B}$\,$\approx$\,15, 
which indicates that NGC~3801 contains a significant dark-matter halo (in agreement with M/L ratios observed in other \HI-rich early-type galaxies; \citealt{ber93}, \citealt*{fra94}, \citealt{mor97}, \citealt{wei08}). Considering only the \HI\ emission, we find M$_{\rm HI}$/$L_{\rm B} = 0.06$ for NGC~3801,\footnote{Note that this estimate does not take into account the \HI\ gas detected in absorption, therefore the true value of M$_{\rm HI}$/$L_{\rm B}$ is somewhat larger.} which is within the range of values observed for other \HI-rich early-type galaxies \citep{oos07,oos10} and polar-ring galaxies \citep{bet01}.

The radio source is aligned roughly along the major axis of the \HI\ disc, indicating that the radio jets are propagating directly into the disc. Perpendicular to the large-scale \HI\ disc and radio axis, \citet{das05} detected a 4.6 kpc wide CO disc, which follows a prominent inner dust-lane \citep{ver99}. We refer to \citet[][their figure 5]{das05} for an excellent illustrative image of the inner region of NGC~3801 (which shows contour plots of the CO(1-0) and VLA 20cm radio continuum overlaid onto an HST image of the inner dust features). 

\citet{das05} also detected a CO clump that is not part of the inner CO disc and aligned along the EW dust-lane, with a projected distance of about 2 kpc from the nucleus. They dismiss the likelihood that this CO clump is part of a second disc, given that it would require the geometry of the inner and outer disc to be extremely rare (roughly one in a thousand), namely such that both discs would have to be edge-on, within 10$^{\circ}$ of the line-of-sight and perpendicular to each other to within 10$^{\circ}$. Our \HI\ results, however, show that NGC~3801 does in fact satisfy the conditions of two perpendicular, edge-on discs (an inner kpc-scale CO disc and a large-scale outer \HI\ disc). Figure \ref{fig:HIdisc} shows that the CO clump detected by \citet{das05} likely represents cold molecular gas that is part of the large-scale outer disc. Given that two discs with intersecting orbits are not stable, it is likely that the large-scale \HI\ disc has a central gap and thus may not continue all the way down to the kpc-scale central region. 

Our \HI\ results (Fig. \ref{fig:HIdisc} -- {\sl middle} and Fig. \ref{fig:HIchanmaps}), however, show that there is neutral gas in the central region, which has a steep velocity gradient relative to the bulk of the gas in the large-scale \HI\ disc. This \HI\ component is most clear in Fig. \ref{fig:HIchanmaps} at v\,=\,3165 \kms\ (panel 1), v\,=\,3251 \kms\ (panel 3) and v\,=\,3728 \kms\ (panel 11), of which the latter two correspond to the velocity of the two clumps of CO gas that represent the central molecular disk in \citet{das05}. This could indicate an \HI\ counterpart to the central CO disc, although the spatial resolution of our \HI\ observations is too poor to verify this.

From an \HI\ perspective, NGC~3801 is not unique. Similar to NGC~3801, examples of other early-type systems with a perpendicular inner and outer disc are known to exist, possibly related to the tri-axial nature of the mass distribution. These include NGC~5266 \citep{mor97} and NGC~2685 \citep[where the two discs/rings are likely a projection effect of an extremely warped disc;][]{joz09}. NGC~3801 also bears resemblance to polar-ring galaxies, such as NGC4650A \citep{arn97} and UGC~9796 \citep*{cox06}.

\begin{table*}
\caption{\HI\ companions}
\vspace{0.4cm}
\label{tab:companions}
\begin{tabular}{clcccc}
$\#$ & Name & D (kpc) & v$_{\rm sys}$ (\kms) & $\Delta$v & M$_{\rm HI}$ (M$_{\odot})$ \\
\hline
1 & NGC~3801 & - & 3552 $\pm$ 20 & 607 & $1.3 \times 10^9$ \\
2 & NGC~3802 & 33  & 3250 $\pm$ 40 & 398 & $3.9 \times 10^7$ \\
3 & J114011.27+174846.6 & 75 & 3674 $\pm$ 20 & 103 & $3.1 \times 10^8$  \\
4 & KUG~1137+179 & 49 & 3858 $\pm$ 40 & 308 & $2.0 \times 10^9$  \\
5 & NGC~3790  & 102 & 3373 $\pm$ 40 & 380 & $6.3 \times 10^8$  \\
6 & NGC~3806 & 118 & 3486 $\pm$ 20 & 180 & $4.0 \times 10^9$  \\
  & `Cloud A' & 83 & 3657 $\pm$ 20 & - & $3.8 \times 10^7$ \\ 
  & `Cloud B' & 132 & 3496 $\pm$ 20 & - & $5.0 \times 10^7$ \\ 
\end{tabular} 
\flushleft 
{Notes -- D is the projected distance to NGC~3801; v$_{\rm sys}$ is the central optical barycentric velocity derived from our \HI\ data; $\Delta$v is the velocity range covered by the \HI\ emission; M$_{\rm HI}$ is the total \HI\ mass detected in emission.}
\end{table*} 

\subsection{Formation history and AGN triggering}
\label{sec:formation}

Based on its peculiar optical morphology and CO content in the inner few kpc, it has been suggested that NGC~3801 could have been involved in a gas-rich galaxy-galaxy merger \citep[e.g.][]{hec86,das05,cro07}. Our \HI\ results are in agreement with such a scenario, given that the formation of a large-scale \HI\ disc can be the natural outcome of a galaxy merger over significant time-scales \citep[e.g.][]{emo06,emo08,ser06}. Simulations of merging galaxies by \citet{mat07} show that gaseous and stellar discs in the progenitor galaxies can be tidally disrupted, depending on the orbital parameters of the merging galaxies (in particular direct encounters can create large gaseous tidal-tails). \citet{bar02} shows that part of this expelled material can be re-accreted back onto the newly formed host galaxy and settle into a large-scale gas disc on time-scales of one to several Gyr. The bulk of the \HI\ gas in the disc of NGC~3801 shows regular rotation, which means that the gas must have had enough time to settle into the observed disc-like structure. As a rough estimate of the merger time-scale, we assume that this process must have lasted at least one orbital period, or $\geq 3.5 \times 10^{8}$ yr.

The age of the radio source is significantly less than this \citep[$\leq 2.4 \times 10^{6}$ yr;][]{cro07}. While there is no evidence for a direct causal connection between the possible merger event and the triggering of the radio source, there would have been a significant time-delay between the start of the merger and the onset of the current episode of the radio-AGN activity. Similar time-delays between merger (and associated starburst episode) and radio-AGN activity have been observed in other nearby radio galaxies \citep[e.g.][]{eva99,tad05,emo06,emo08,lab08}.

We note, however, that our \HI\ results alone do not provide conclusive evidence that the large-scale \HI\ disc in NGC~3801 was formed as a result of a major merger. Minor mergers or interactions with satellite galaxies (similar to the ongoing interaction between NGC~3801 and NGC~3802) may also have played a role in both shaping the \HI\ disc as well as triggering nuclear activity in NGC~3801. Given that the bulk of the \HI\ gas that we find in our data cube is associated with companion galaxies that are all within $v \sim 300$ \kms\ from the central velocity of NGC~3801, it is possible that the \HI\ gas in the disc of NGC~3801 was delivered by former gas-rich companions over the course of several Gyr (similar to what is discussed by \citealt{str10_NGC1167} for the large \HI\ disc in radio galaxy NGC\,1167).

Another possibility is that accretion of cold gas played a major role in the formation of the \HI\ disc. Simulations by \citet{mac06} show that extended rings of gas in polar-ring galaxies may form naturally through the accretion of cold gas falling in along large filamentary structures, while \citet{bou03} argue that it is more likely for polar-ring galaxies to acquire their gaseous discs through tidal accretion from a gas-rich donor galaxy rather than a violent merger. A similar mechanism may have formed the large-scale \HI\ disc in NGC~3801.

\subsection{Radio source evolution}

As mentioned above, the alignment of the small radio source in NGC~3801 with the large-scale outer \HI\ disc makes NGC~3801 an ideal nearby radio galaxy for studying the early stage of radio-source evolution and its interaction with the surrounding cold ISM.


\citet{cro07} find evidence for shocked heated shells of hot gas associated with the radio source in NGC~3801, caused by supersonic, over-pressured radio jets. Their findings indicate that the radio source in NGC~3801 is heavily interacting with the surrounding ISM. In proceedings based on high-resolution VLA \HI\ observations, \citet{hot09} report the presence of a broad, faint, blueshifted absorption and an \HI\ clump associated with the shocked shell around the eastern lobe, possibly reflecting a gas outflow. Our lower resolution \HI\ data do not show clear evidence for a broad, blueshifted absorption, but such a faint \HI\ absorption feature could be filled in with \HI\ emission within the relatively large beam of our observations. We do detect an apparent blueshifted narrow absorption component against the western radio lobe, but as discussed in Sect. \ref{sec:absorption}, this most likely represents \HI\ gas as part of the large-scale \HI\ disc. 

Similar to NGC~3801, there are other known radio galaxies in which the radio source propagates into an outer \HI\ disc. This includes Coma~A, where the radio lobes ionise part of the disc \citep[][]{mor02} and IC~5063, where the radio jets produce fast outflows of \HI\ gas \citep[][]{mor98,oos00}. Particularly interesting is also the similarity with the disc-dominated radio galaxy B2~0722+30, which hosts a small \FRI\ radio source (similar to NGC~3801) that propagates close to the plane of the disc, in the direction of a heavily interacting \HI-rich galaxy pair \citep{emo09}.

\subsection{\HI-rich radio galaxies: the general picture}
\label{sec:general}

\begin{figure}
\centering
\includegraphics[width=0.45\textwidth]{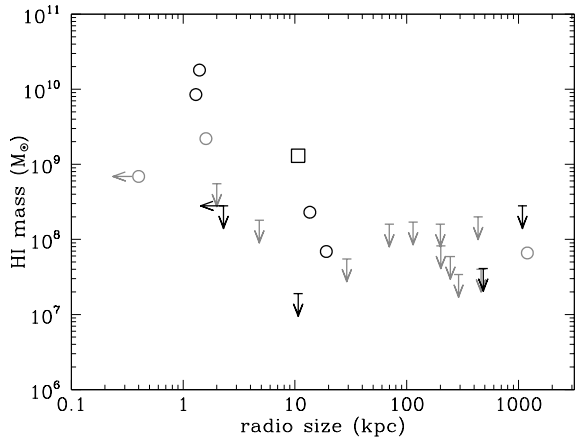}
\caption{Adapted from \citet{emo10}. Shown is the \HI\ mass content (or upper limit) plotted against the total linear extent of the radio source for a complete sample of nearby non-cluster, low-power (\FRI-type) radio galaxies. The square represents NGC~3801. See \citet{emo07} and \citet{emo10} for details. For completeness, radio host galaxies designated as elliptical (E) in the NASA Extra-galactic Database (NED) are plotted in grey, while non-ellipticals (S0 and merger systems) are black. While this reflects the long known idea that the relative \HI\ fraction seems to rise from ellipticals to S0 and later-type systems \citep[e.g.][]{war86}, it also shows that the \HI-rich compact radio sources in this sample are hosted by galaxies with a variety of optical morphologies \citep[see][for detailed optical imaging]{emo10}.}
\label{fig:HImasssize}
\end{figure}

In general, roughly 10$\%$ of early-type galaxies outside clusters contain large amounts of \HI\ gas (M$_{\rm HI} \geq 10^{9}$\,M$_{\odot}$), often distributed in large-scale disc- or ring-like structures \citep{sad01,mor06,oos07,oos10}.\footnote{\citet{dis07} and \citet{oos10} show that the \HI\ detection rate of early-type galaxies in the Virgo Cluster is dramatically lower.} In \citet{emo10}, we showed that the overall \HI\ properties of nearby, non-cluster low-power radio galaxies are similar to those of radio-quiet early-type galaxies, with large-scale \HI\ discs detected in a significant fraction of radio-loud systems. In addition, we showed that there is a clear segregation in \HI\ content with radio size, in the sense that massive, large-scale \HI\ discs are only found in the host galaxies of compact radio sources, while none of the more extended \FRI\ radio sources studied contain similar amounts of \HI\ gas \citep{emo07,emo10}. The radio source in NGC~3801 is $\sim 11$ kpc in diameter \citep{jen82}. As illustrated in Fig. \ref{fig:HImasssize}, the \HI\ properties of NGC~3801 are thus similar to those of a significant fraction of nearby low-power compact radio sources. 

It is likely that in these systems the early evolution of the radio source is intimately linked with the surrounding cold ISM, and that either confinement by the dense ISM or inefficient fuelling (through clumpiness of the cold ISM and associated non-steady accretion) keeps the radio source compact \citep[see discussions in][]{emo07,emo10}. An alternative scenario, recently raised by \citet{mor11} in a study of the powerful compact radio source PKS~1814-637, suggests that the radio emission in these \HI-rich compact radio sources may be boosted, at least temporarily, by the interaction of the radio jets with the rich ISM, implying that intrinsically these compact sources may reflect a link to radio-AGN in Seyfert galaxies (see also \citealt{emo10}, where we speculate that the transition from radio-AGN in Seyferts to the brighter compact radio sources to classical \FRI s may perhaps be linked to a transition from well-defined star-forming \HI\ discs to low surface brightness \HI\ discs to \HI-poor host galaxies).

\subsubsection{NGC~3801: a classical \FRI?}
\label{sec:classical}

NGC~3801 is classified as an \FRI\ radio galaxy, based on its total radio power ($P_{\rm 1.4\,GHz} = 2.9 \times 10^{23}$\,W\,Hz$^{-1}$), radio source morphology \citep{jen82,cro07} and weak emission-line spectrum \citep{hec86}. However, X-ray observations by \citet{cro07} show that NGC~3801 has supersonic, over-pressured radio jets, not commonly found in \FRI\ sources \citep{cro08}. These X-ray studies also reveal a high absorbing column towards the nucleus (possibly representing an obscuring torus), something that is generally not associated with \FRI\ sources \citep{har06,cro07}. \citet{cro07} suggest that these properties may more closely resemble those of high-power, high-excitation \FRII\ sources, which are believed to be post-merger systems that are fuelled by cold gas driven into the central region to form a radiatively-efficient classical accretion disc (contrary to low-power/low-excitation \FRI s, which are believed to be fuelled mainly through a quasi-spherical Bondi accretion of circum-galactic hot gas directly onto the nucleus; see e.g. \citealt{hec86}, \citealt{bau92}, \citealt{har06}, \citealt*{har07}, \citealt{bal08}, \citealt{emo10}, \citealt{ram11}). This scenario is in agreement with our \HI\ results, which show that NGC~3801 is likely a post-merger system with large amounts of cold gas.\footnote{Our \HI\ results do not show evidence that cold gas is being accreted onto the central region (see Section \ref{sec:absorption}), but note that the accretion rate necessary to sustain AGN activity could be so low that it is well beyond our observing capabilities \citep{gor89}.}

However, as we saw in Sect. \ref{sec:general}, the presence of a large-scale \HI\ disc is not unusual for low-power compact radio sources like NGC~3801. We speculate that the possible link to radio-loud Seyferts (discussed in Sect. \ref{sec:general}) could perhaps also explain some of the unusual characteristics of NGC~3801. Many Seyfert-AGN also show heavy obscuration in X-rays (e.g. \citealt{mai98}, \citealt*{ris99}, \citealt{cap06}) and can be found in disc-dominated, interacting galaxies that contain large amounts of \HI\ gas \citep[e.g.][]{kuo08}.
 
Regardless of the exact nature of the radio-loud AGN, the combined X-ray and \HI\ results suggests that NGC~3801 is a system with over-pressured radio jets that are embedded in (and interacting with) a rich, ambient cold ISM that has been deposited after a gas-rich merger; features not or no longer present in more evolved \FRI\ radio sources. As discussed by \citet{cro07}, this resembles the case of Centaurus~A (see also \citealt{kra03,gor90,sch94,str10}). X-ray observations of other \HI-rich compact \FRI\ radio sources may reveal how unique NGC~3801 (and Cen~A) are in the evolution of classical radio sources.

\section{Conclusions}
\label{sec:conclusions}

We have shown that the \HI\ gas in NGC~3801 -- detected in earlier single-dish observations by \citet{hec83} and \citet{dup96} -- is distributed in a regularly rotating large-scale \HI\ disc. We also detect \HI\ gas in the close companion NGC~3802 (which is in apparent ongoing interaction with NGC~3801) as well as several other galaxies and gas clouds in the immediate environment of NGC~3801. The small radio source in NGC~3801 is propagating directly into the large-scale \HI\ disc, making NGC~3801 an important system for investigating in detail the evolution of a small radio source through its host galaxy's cold ISM. Because of the similarity between the \HI\ properties of NGC~3801 and those of other nearby low-power compact radio sources, studying NGC~3801 may reveal important insight into the early evolution of at least a significant fraction of low-power radio sources.

\section*{Acknowledgments}
We thank the referee Dhruba Saikia for suggestions that significantly improved this paper. The Westerbork Synthesis Radio Telescope is operated by the ASTRON (Netherlands Institute for Radio Astronomy) with support from the Netherlands Foundation for Scientific Research (NWO).


\end{document}